\documentclass[conference]{IEEEtran}

\IEEEoverridecommandlockouts

\usepackage{cite}
\usepackage{amsmath,amssymb,amsfonts}
\usepackage{graphicx}
\usepackage{textcomp}
\usepackage{xcolor}
\usepackage{comment}
\usepackage{booktabs}
\usepackage{multirow}
\usepackage{array}
\usepackage{tcolorbox}
\usepackage{balance}
\usepackage{microtype}

\def\BibTeX{{\rm B\kern-.05em{\sc i\kern-.025em b}\kern-.08em
    T\kern-.1667em\lower.7ex\hbox{E}\kern-.125emX}}

\begin{document}

\title{SENTINEL: A Multi-Pathway Architecture for Detecting Living-Off-the-Land APT Attacks on Windows Command Lines}




\author{
  \IEEEauthorblockN{
    Ahad Bin Islam Shoeb\IEEEauthorrefmark{1},
    Kamrul Hasan\IEEEauthorrefmark{2},
    Jamal Uddin Tanvin\IEEEauthorrefmark{3},
    Liang Hong\IEEEauthorrefmark{2},
    Imtiaz Ahmed\IEEEauthorrefmark{5},\\
    Md Arif Billah\IEEEauthorrefmark{1}, and
    Al Amin\IEEEauthorrefmark{4}
  }
  \IEEEauthorblockA{
    \IEEEauthorrefmark{1}University of Dhaka, Dhaka, Bangladesh, 
    \IEEEauthorrefmark{2}Tennessee State University, Nashville, TN, USA\\
    \IEEEauthorrefmark{3}Military Institute of Science and Technology, Dhaka, Bangladesh, 
    \IEEEauthorrefmark{4}Huston--Tillotson University, Austin, TX, USA\\
    \IEEEauthorrefmark{5}Howard University, Washington, DC, USA\\
    E-mail: ahadbinislam-2019117810@cs.du.ac.bd, mhasan1@tnstate.edu, jamaluddintanvin@outlook.com, lhong@tnstate.edu,\\ 
    imtiaz.ahmed@howard.edu, mdarif-2019617815@cs.du.ac.bd, and aamin@htu.edu 
  }
}

\maketitle

\begingroup
  \renewcommand\thefootnote{}\footnote{%
    Accepted for publication in the Proceedings of the 2026 IEEE Military Communications Conference (MILCOM 2026). This is the authors' accepted version; the final published version will appear in IEEE Xplore. \copyright~2026 IEEE. Personal use of this material is permitted. Permission from IEEE must be obtained for all other uses, in any current or future media, including reprinting/republishing this material for advertising or promotional purposes, creating new collective works, for resale or redistribution to servers or lists, or reuse of any copyrighted component of this work in other works.%
  }%
  \addtocounter{footnote}{-1}%
\endgroup

\begin{abstract}
Living-Off-the-Land (LOTL) is the dominant evasion technique of Advanced Persistent Threat (APT) actors, exploiting legitimate Windows utilities to conduct malicious operations without deploying custom malware and enabling state-sponsored campaigns to maintain persistent access within military and critical defense infrastructure for extended periods. Existing detection methods fail against obfuscated commands and multi-stage attack sequences, as demonstrated by the Volt Typhoon APT campaign, which maintained undetected access to U.S. critical infrastructure for over 18 months using exclusively signed Windows utilities. We present SENTINEL, a multi-pathway architecture integrating BERT-based semantic encoding, character-level CNN for obfuscation invariance, inter-command attention for multi-stage pattern recognition, and autoencoder-based anomaly scoring. Evaluated on a balanced Volt Typhoon benchmark derived from Microsoft and CISA threat intelligence advisories, SENTINEL achieves 92.0\% accuracy on documented state-sponsored attack commands and 91.2\% on obfuscated variants, compared to 74.0\% and 72.0\% for standalone BERT. Per-class analysis reveals that models achieving over 98\% overall validation accuracy on imbalanced data exhibit only 44--58\% malicious recall on balanced adversarial sets. Character-level processing contributes 5.6 percentage points of obfuscation invariance, and the 8.0 percentage point gap over augmentation-only baselines confirms structural architectural value beyond data-driven robustness alone.
\end{abstract}


\section{Introduction}
Living-Off-the-Land (LOTL) has emerged as the primary evasion technique of Advanced Persistent Threat (APT) actors. Rather than deploying custom malware, APT operators exploit trusted Windows utilities, including PowerShell, WMI, \texttt{certutil}, and \texttt{schtasks} to conduct malicious operations while blending with routine administrative activity~\cite{cisa_lotl_guidance_2024,barr2021survivalism}. Because these binaries are whitelisted by enterprise security controls and used identically by system administrators and adversaries, LOTL-based APT attacks are structurally resistant to signature-based detection, binary inspection, and sandbox analysis. The attack surface is not a novel executable but the administrative tooling already present in the target environment.

The strategic severity of this threat for military and defense networks was established in May 2023, when Microsoft Threat Intelligence documented the Volt Typhoon campaign: a PRC state-sponsored operation that compromised U.S. critical infrastructure across the communications, energy, and water sectors while evading commercial endpoint detection and response (EDR) products for over 18 months~\cite{microsoft2023volttyphoon}. Palo Alto Networks Unit 42 corroborated the campaign scope and tactical progression~\cite{unit422024volttyphoon}, while joint CISA/NSA/FBI advisories identified systematic obfuscation and multi-stage command sequencing as the primary evasion mechanisms~\cite{cisa_lotl_guidance_2024}. For defense networks where the same Windows administrative utilities underpin both legitimate operations and adversary tradecraft, LOTL represents a paradigm shift requiring detection capabilities well beyond what current solutions provide.

Existing detection approaches fail against sophisticated LOTL attacks for three structural reasons. \textbf{First,} token-based methods including TF-IDF and BERT~\cite{devlin2019bert} fragment obfuscated commands through base64 encoding, string concatenation, and alias substitution into non-informative subwords, causing accuracy to drop from over 98\% on clean validation data to 52--74\% on adversarial inputs. \textbf{Second,} single-command classifiers treat each command independently and cannot recognize that individually benign commands such as \texttt{whoami}, \texttt{net user}, \texttt{reg query}, and \texttt{ntdsutil} constitute an attack progression when executed in a coordinated sequence, mirroring Volt Typhoon's reconnaissance-to-credential-access workflow~\cite{cisa_lotl_guidance_2024}. \textbf{Third,} supervised classifiers trained on known patterns often produce highly confident misclassifications for novel tool combinations and encoding strategies absent from the training distribution, thereby providing no effective triage signal for zero-day LOTL techniques.~\cite{mirsky2018kitsune}.

We present SENTINEL (\textbf{S}emantic \textbf{EN}coding with charac\textbf{T}er-level \textbf{IN}variance and int\textbf{E}r-command modeling for \textbf{L}OTL detection), the first end-to-end architecture to jointly address all three failure modes through four complementary detection pathways: (1) a BERT-based semantic encoder~\cite{devlin2019bert} for contextual command intent modeling; (2) a character-level CNN~\cite{zhang2015character} for obfuscation-invariant surface analysis; (3) an inter-command attention module for multi-stage attack correlation; and (4) an autoencoder-based anomaly detector for distributional shift triage.

The contributions of this work are as follows:
\begin{enumerate}
\item \textbf{A unified multi-pathway detector.} We design SENTINEL, to our knowledge the first single end-to-end trainable model that simultaneously confronts obfuscation invariance, semantic intent discrimination, multi-stage temporal correlation, and distributional-shift triage for Windows command-line LOTL detection, rather than addressing these failure modes in isolation.
\item \textbf{An intelligence-grounded balanced benchmark.} We curate a class-balanced Volt Typhoon evaluation set from government and vendor threat-intelligence advisories and use it to drive a per-class analysis that exposes malicious-recall failures which conventional imbalanced metrics conceal.
\item \textbf{An evaluation-methodology finding with operational stakes.} We demonstrate that headline validation accuracy exceeding 98\% can co-occur with malicious recall as low as 44--58\% on balanced adversarial inputs, showing that standard reporting practice systematically overstates readiness for defense-network monitoring.
\item \textbf{Quantified architectural (not data-driven) gains.} Through leave-one-out ablation, we isolate an 8.0 percentage-point improvement attributable to multi-pathway integration beyond augmentation-only baselines, establishing that the robustness is structural and cannot be recovered by training-data diversity alone.
\end{enumerate}

\section{Background and Related Work}
\subsection{Traditional Machine Learning Approaches}

Early LOTL detection relied on handcrafted features with classical classifiers. Hendler et al.~\cite{hendler2018detecting} achieve 91--94\% F1 on PowerShell detection using TF-IDF with Random Forest; however, any syntactic transformation defeats lexical feature extraction~\cite{li2019effective}. Bag-of-words and n-gram representations cannot model cross-command dependencies or recognize semantic equivalence between utilities serving identical attack purposes, such as \texttt{ntdsutil} and \texttt{vssadmin} for credential access. These approaches share a structural limitation: they operate on surface form rather than semantic intent, making them unsuitable for LOTL scenarios where defenders and adversaries share the same vocabulary of legitimate tools~\cite{barr2021survivalism}.

\subsection{Deep Learning and Transformer Methods}

LSTM-based models provide temporal sequence modeling but suffer gradient vanishing over long command chains~\cite{du2017deeplog}, losing early reconnaissance context before credential-theft commands appear. Character-level CNNs~\cite{zhang2015character,kim2014convolutional} handle obfuscation by operating below token boundaries but cannot distinguish malicious from benign commands with identical character distributions (e.g., \texttt{ntdsutil snapshot} for disaster recovery versus credential harvesting). BERT~\cite{devlin2019bert} advances semantic understanding but was designed for natural language: WordPiece tokenization fragments base64-encoded payloads into meaningless subwords, causing a 25--27 percentage point accuracy drop on adversarial inputs. Domain-adapted variants such as SecurityBERT~\cite{aghaei2022securebert} and LogBERT~\cite{guo2021logbert} improve security-event modeling but inherit the same tokenization vulnerability and do not provide the character-level obfuscation invariance that command-line LOTL detection requires.

\subsection{Specialized LOTL Systems}

The LOLBAS project~\cite{lolbas2024} catalogs 347+ abusable Windows binaries, but signature matching breaks under any syntactic transformation. LOLWTC~\cite{ding2023lolwtc} reaches 93.83\% validation accuracy via handcrafted features yet collapses to 52.0\% on Volt Typhoon and 48\% on obfuscated variants once encoding destroys its feature basis. PowerPeeler~\cite{li2024powerpeeler} robustly handles PowerShell-specific obfuscation but offers no coverage for WMI, \texttt{certutil}, or \texttt{ntdsutil} utilities central to Volt Typhoon activity. Ongun et al.~\cite{ongun2021living} reduce annotation cost through active learning, and CmdCaliper~\cite{huang2024cmdcaliper} shows that embedding quality is critical for downstream detection, but both rely on command-structure features that degrade under obfuscation. Critically, no prior architecture jointly addresses obfuscation invariance, semantic discrimination, temporal sequence modeling, and anomaly detection in a single end-to-end trainable framework evaluated against documented state-sponsored campaigns.

\section{SENTINEL Architecture}
\subsection{Overview}

\begin{figure*}[!htb]
\centering
\includegraphics[width=0.9\textwidth,keepaspectratio]{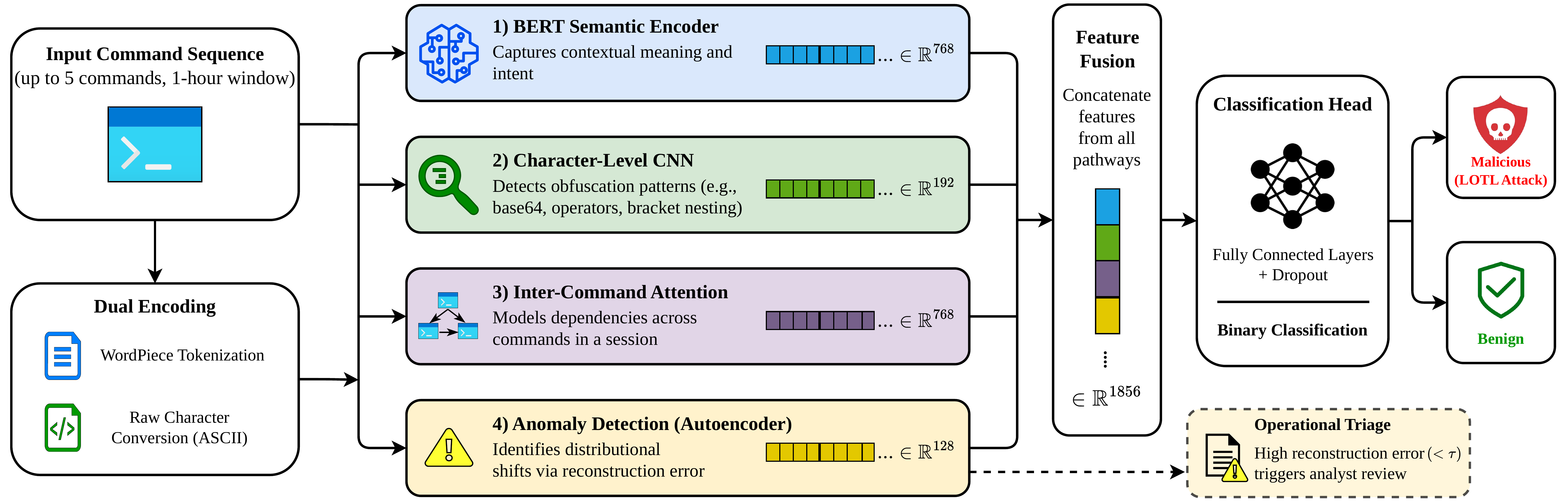}
\caption{SENTINEL architecture: SENTINEL encodes inputs using WordPiece and characters, processes them through BERT, sequence attention, CharCNN, and anomaly detection, and then fuses features into a 1856-D vector for malicious/benign classification.}
\label{fig:architecture}
\end{figure*}

SENTINEL detects LOTL attacks through four parallel pathways integrated via learned fusion. Fig.~\ref{fig:architecture} illustrates the full architecture: input commands flow through dual encoding (WordPiece tokenization and raw character conversion) into four independent pathways whose outputs are concatenated into a unified 1856-dimensional representation before the classification head. 
The BERT pathway captures semantic intent through contextual embeddings, recognizing credential-access commands such as \texttt{reg save HKLM\textbackslash SAM} despite syntactic variation. The CharCNN pathway provides obfuscation invariance by operating on raw character sequences, detecting base64 padding, operator manipulation, and bracket nesting independent of tokenization. The inter-command attention pathway models multi-stage attack progressions by learning that reconnaissance commands followed by credential access constitute coordinated campaigns. The anomaly detection pathway flags distributional shifts through autoencoder reconstruction error, identifying novel command combinations absent from training data.

\subsection{BERT Semantic Encoder}

The BERT encoder produces contextual embeddings capturing command intent. Each token $t_i$ embeds as:
\begin{equation}
\mathbf{e}_i = \mathbf{W}_{\text{token}}[t_i] + \mathbf{W}_{\text{pos}}[i] + \mathbf{W}_{\text{seg}}[s]
\label{eq:bert_embed}
\end{equation}
where $\mathbf{W}_{\text{token}} \in \mathbb{R}^{V \times 768}$ provides token embeddings, $\mathbf{W}_{\text{pos}}$ encodes absolute positions, and $\mathbf{W}_{\text{seg}}$ encodes segment identifiers. Twelve transformer layers~\cite{vaswani2017attention} apply multi-head self-attention and feedforward transformations, producing output $\mathbf{H}^{(12)} = [\mathbf{h}_1, \ldots, \mathbf{h}_n]$ with $\mathbf{h}_i \in \mathbb{R}^{768}$ flowing to three downstream pathways: inter-command attention, anomaly detection, and feature fusion.

\subsection{Character-Level CNN}

The CharCNN pathway detects obfuscation patterns invisible to token-based methods. Commands convert to fixed-length ASCII sequences $\mathbf{x}_{\text{char}} \in \mathbb{Z}^{512}_{[0,255]}$. Each character embeds through $\mathbf{W}_{\text{char}} \in \mathbb{R}^{256 \times 32}$. Three parallel convolutions with filter sizes $k \in \{3, 4, 5\}$ extract character $n$-grams:
\begin{equation}
\mathbf{f}^{(k)}_i = \text{ReLU}\!\left(\sum_{j=0}^{k-1} \mathbf{W}^{(k)}_j\, \mathbf{e}_{i+j} + b^{(k)}\right)
\label{eq:conv}
\end{equation}
Global max-pooling identifies the strongest pattern per filter size: $\mathbf{g}^{(k)} = \max_i \mathbf{f}^{(k)}_i \in \mathbb{R}^{64}$. Concatenation yields $\mathbf{h}_{\text{char}} = [\mathbf{g}^{(3)}; \mathbf{g}^{(4)}; \mathbf{g}^{(5)}] \in \mathbb{R}^{192}$, capturing base64 padding (\texttt{==}), PowerShell concatenation operators, and bracket nesting regardless of position within the command string.

\subsection{Inter-Command Attention}

This module models dependencies across commands within a session. Each command $c_i$ in sequence $S = [c_1, \ldots, c_T]$ with $T \leq 5$, grouped in 1-hour windows, is encoded via BERT and mean-pooled to $\mathbf{h}_i \in \mathbb{R}^{768}$. Multi-head attention with $H = 4$ heads and $d_k = 192$ computes inter-command dependencies:
\begin{equation}
\text{Attn}(\mathbf{Q}, \mathbf{K}, \mathbf{V}) = \text{softmax}\!\left(\frac{\mathbf{Q}\mathbf{K}^{T}}{\sqrt{d_k}}\right)\mathbf{V}
\label{eq:attn}
\end{equation}
Training applies dual supervision balancing single-command and sequence-level classification:
\begin{equation}
\mathcal{L}_{\text{seq}} = \alpha\,\mathcal{L}_{\text{single}} + (1-\alpha)\,\mathcal{L}_{\text{sequence}}, \quad \alpha = 0.6
\label{eq:seq_loss}
\end{equation}
Training sequences ($n = 1{,}247$) include 892 benign administrative workflows and 355 malicious progressions labeled according to MITRE ATT\&CK phases~\cite{mitre2024attack}, with inter-rater agreement of Cohen's $\kappa = 0.87$.

\subsection{Anomaly Detection}

The anomaly branch identifies out-of-distribution commands through autoencoder reconstruction error~\cite{mirsky2018kitsune}. The autoencoder is pretrained solely on benign training data and frozen during joint SENTINEL training, preventing data leakage. The encoder compresses $\mathbf{h}_{\text{BERT}} \in \mathbb{R}^{768}$ to a 128-dimensional latent representation $\mathbf{z}_{\text{anomaly}}$ through two linear layers with ReLU activations. A symmetric decoder reconstructs the BERT embedding $\hat{\mathbf{h}}_{\text{BERT}}$. Reconstruction error $\mathcal{L}_{\text{recon}} = \|\mathbf{h}_{\text{BERT}} - \hat{\mathbf{h}}_{\text{BERT}}\|_2^2$ exceeding threshold $\tau = 0.11$ (95th percentile on benign validation data) triggers analyst triage. Ablation confirms 0.4--0.6\% classification impact, positioning this pathway as operational triage rather than primary detection.

\subsection{Feature Fusion and Training}

Feature concatenation forms a unified representation:
\begin{equation}
\mathbf{h}_{\text{concat}} = [\mathbf{h}_{\text{BERT}};\,\mathbf{h}_{\text{char}};\,\mathbf{h}_{\text{temporal}};\,\mathbf{z}_{\text{anomaly}}] \in \mathbb{R}^{1856}
\label{eq:fusion}
\end{equation}
where $1856 = 768 + 192 + 768 + 128$. A fusion layer applies LayerNorm, ReLU, and dropout ($p = 0.3$), followed by two dense layers for final binary classification. SENTINEL optimizes a combined objective:
\begin{equation}
\mathcal{L}_{\text{total}} = \mathcal{L}_{\text{CE}} + \lambda\,\mathcal{L}_{\text{recon}}, \quad \lambda = 0.1
\label{eq:total_loss}
\end{equation}
where $\mathcal{L}_{\text{CE}}$ uses class weights ($w_{\text{malicious}} = 8.08$, $w_{\text{benign}} = 0.53$) to address the 15:1 training imbalance. Training uses the AdamW optimizer with learning rate $2\times10^{-5}$, batch size 32, and early stopping with patience 3. Performance is stable across 5 random seeds: Volt Typhoon $92.0\% \pm 1.2\%$, obfuscated $91.2\% \pm 0.8\%$.

\section{Experimental Evaluation}
\subsection{Dataset and Setup}

Our dataset contains 7,049 commands: 6,613 benign and 436 malicious. Benign samples originate from enterprise Sysmon logs (3,847 commands from 50 workstations over 6 months), GitHub administrative scripts (1,523), and automated system tasks (1,243). Malicious samples include commands from documented APT29, APT32, Lazarus, and Turla campaigns (187)~\cite{saha2025expert}, red-team exercises using Empire and Cobalt Strike (134), and LOLBAS abuse cases (115). Data was split 70/15/15 (training: 4,934; validation: 1,058; testing: 1,057), with SHA-256 deduplication removing 312 exact duplicates and Levenshtein filtering ($>$90\% overlap threshold) removing 89 near-duplicates. All Volt Typhoon samples were held out completely from training.

\textit{Volt Typhoon Evaluation Set} ($n = 100$, balanced). Fifty malicious commands were manually extracted from Microsoft Threat Intelligence documentation~\cite{microsoft2023volttyphoon} and CISA advisories~\cite{cisa_lotl_guidance_2024} spanning eight attack phases: SSH key enumeration (6), browser profile discovery (4), remote access tool queries (12), credential harvesting (8), network reconnaissance (7), persistence (5), lateral movement (5), and exfiltration (3). Fifty benign commands using the same utilities (\texttt{reg query}, \texttt{dir}, \texttt{ntdsutil}) in legitimate administrative contexts were sourced from enterprise Sysmon logs matched by utility type and argument count. Two security analysts achieved Cohen's $\kappa = 0.91$ inter-rater agreement.

\textit{Obfuscated Dataset} ($n = 250$). Generated from test-split commands using eight deterministic transformation techniques: base64 encoding (40), string concatenation (35), backtick injection (30), alias substitution (35), case randomization (30), character encoding (25), format operators (30), and whitespace insertion (25).

Experiments used PyTorch 2.0.1, \texttt{bert-base-uncased}, and an NVIDIA Tesla T4 GPU. Baselines span classical ML (BoW, TF-IDF, Random Forest, XGBoost), deep learning (Char-CNN~\cite{zhang2015character}, Token-Char Fusion~\cite{hendler2018detecting}, BERT~\cite{devlin2019bert}, BERT with obfuscation-aware augmentation), and specialized systems (PowerPeeler~\cite{li2024powerpeeler}, LOLWTC~\cite{ding2023lolwtc}).

\subsection{Comparative Results}

\begin{figure}[!htb]
\centering
\includegraphics[width=0.99\columnwidth]{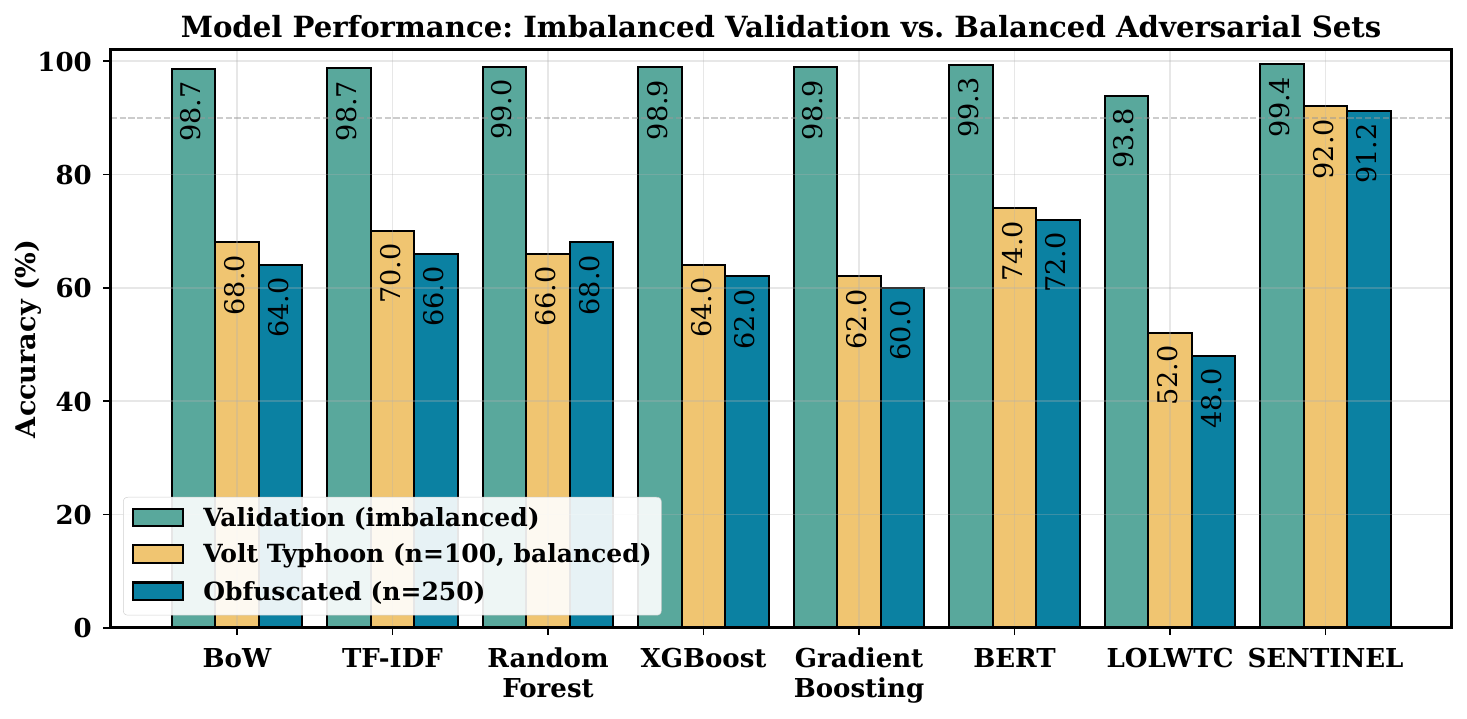}
\caption{SENTINEL shows strong and stable accuracy on validation, Volt Typhoon, and obfuscated tests, while baseline models drop sharply under adversarial conditions.}
\label{fig:comparison}
\end{figure}

Table~\ref{tab:comparison} presents accuracy across all three evaluation sets. On imbalanced validation data, most models exceed 98\% accuracy. The critical finding emerges on adversarial datasets: SENTINEL achieves 92.0\% on Volt Typhoon and 91.2\% on obfuscated inputs, while BERT drops to 74.0\% and 72.0\%, and LOLWTC collapses to 52.0\% and 48.0\%. The 8.0 percentage point gap over BERT with augmentation (84.0\%) on Volt Typhoon confirms that multi-pathway architectural integration provides robustness beyond data augmentation alone. Augmentation improves over vanilla BERT by 10.0 points, while the architecture contributes an additional 8.0 points that data diversity cannot recover. Fig.~\ref{fig:comparison} visualizes accuracy across all three evaluation sets; the dramatic drop in baseline bars from the Validation to Volt Typhoon and Obfuscated columns confirms that the performance degradation is structural, not a tuning artifact.
\begin{table}[!htb]
\renewcommand{\arraystretch}{0.85}
\centering
\caption{Accuracy across all evaluation datasets. }
\label{tab:comparison}
\resizebox{\columnwidth}{!}{%
\begin{tabular}{@{} l c c c @{}}
\toprule
\textbf{Model} & \textbf{Validation} & \textbf{Volt Typhoon} & \textbf{Obfuscated} \\
\midrule
TF-IDF & 98.72\% & 70.0\% & 66.0\% \\
Char-CNN \cite{zhang2015character} & 96.2\% & 78.0\% & 82.4\% \\
Token-Char Fusion \cite{hendler2018detecting} & 97.4\% & 82.0\% & 85.2\% \\
BERT \cite{devlin2019bert} & 99.29\% & 74.0\% & 72.0\% \\
BERT + Augmentation & 99.15\% & 84.0\% & 86.0\% \\
PowerPeeler \cite{li2024powerpeeler} & 97.8\% & 81.0\% & 85.6\% \\
LOLWTC \cite{ding2023lolwtc} & 93.83\% & 52.0\% & 48.0\% \\
\midrule
\textbf{SENTINEL} & \textbf{99.43\%} & \textbf{92.0\%} & \textbf{91.2\%} \\
\bottomrule
\end{tabular}%
}
\begin{minipage}{1.0\columnwidth}
\vspace{0.3em}
\scriptsize \raggedright Volt Typhoon ($n=100$) is balanced (50 benign + 50 malicious); validation is imbalanced (15:1). Validation accuracy masks per-class failures, as revealed in Fig.~\ref{fig:perclass}.
\end{minipage} 
\end{table}


\subsection{Structural Failure Analysis}

\begin{figure}[!htb]
\centering
\vspace{-0.8em}
\includegraphics[width=0.78\columnwidth]{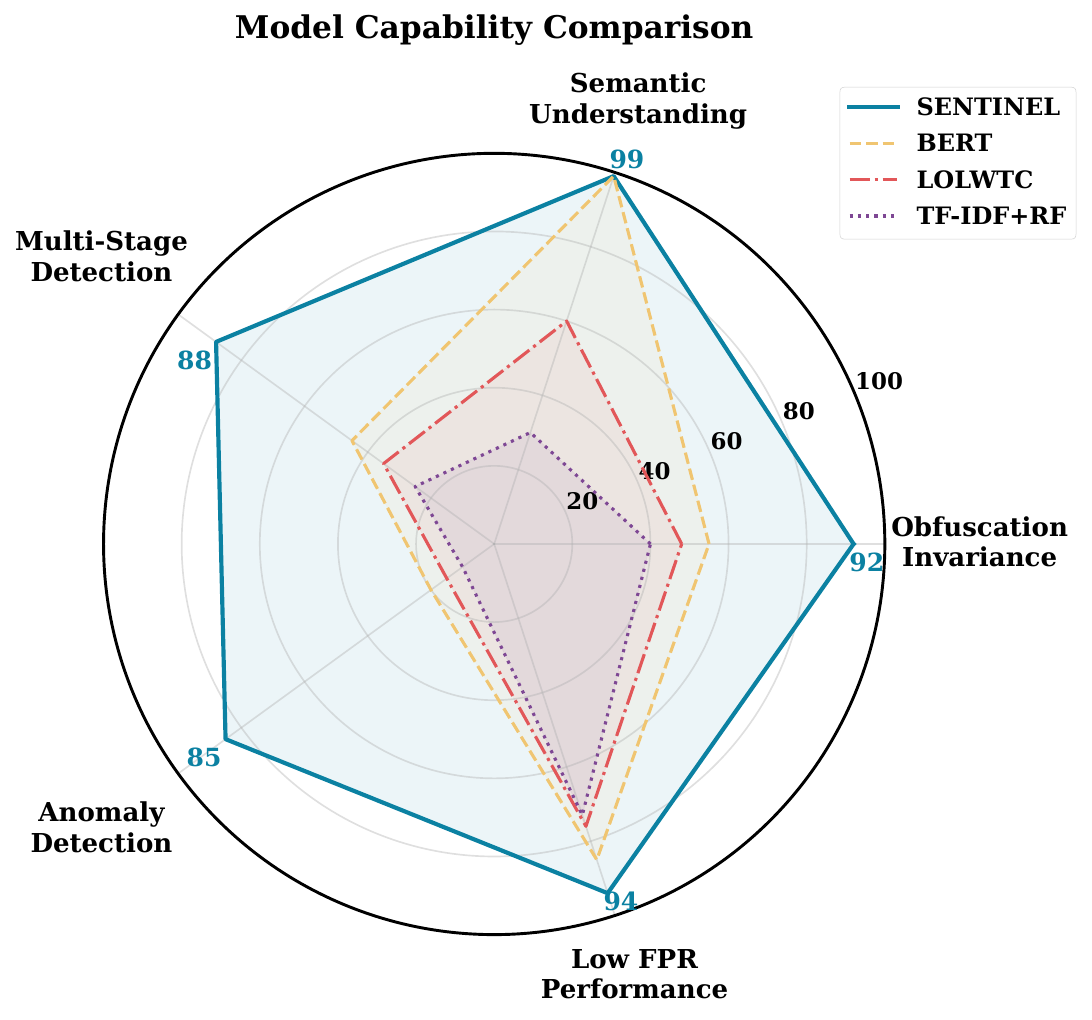}
\caption{Radar chart comparing SENTINEL against baselines across structural challenges.}
\label{fig:structural}
\end{figure}

Fig.~\ref{fig:structural} shows a radar-based evaluation across five architectural capability dimensions derived from the APT threat model, where each axis isolates a specific weakness and the overall area reflects structural robustness. These limitations arise from model design rather than data or tuning constraints.

\textbf{Obfuscation Invariance}: Lexical models (TF-IDF, LOLWTC) break under syntactic changes, while BERT's subword tokenization fails on heavily obfuscated or encoded commands. In contrast, SENTINEL's CharCNN operates at the character level, capturing robust patterns independent of tokenization.

\textbf{Multi-Stage Recognition and Temporal Dispersion}: Independent classifiers miss coordinated attack sequences, and LSTM struggles with long-range dependencies. SENTINEL's inter-command attention models full sequence relationships, enabling stronger multi-stage understanding.

\textbf{Distributional Shift}: Standard classifiers misclassify unseen attack types without uncertainty modeling. SENTINEL's anomaly detection module flags out-of-distribution inputs via reconstruction error, improving robustness to novel techniques.

\subsection{Per-Class Analysis}
\begin{figure}[!htb]
\centering
\vspace{-0.8em}
\includegraphics[width=0.99\columnwidth]{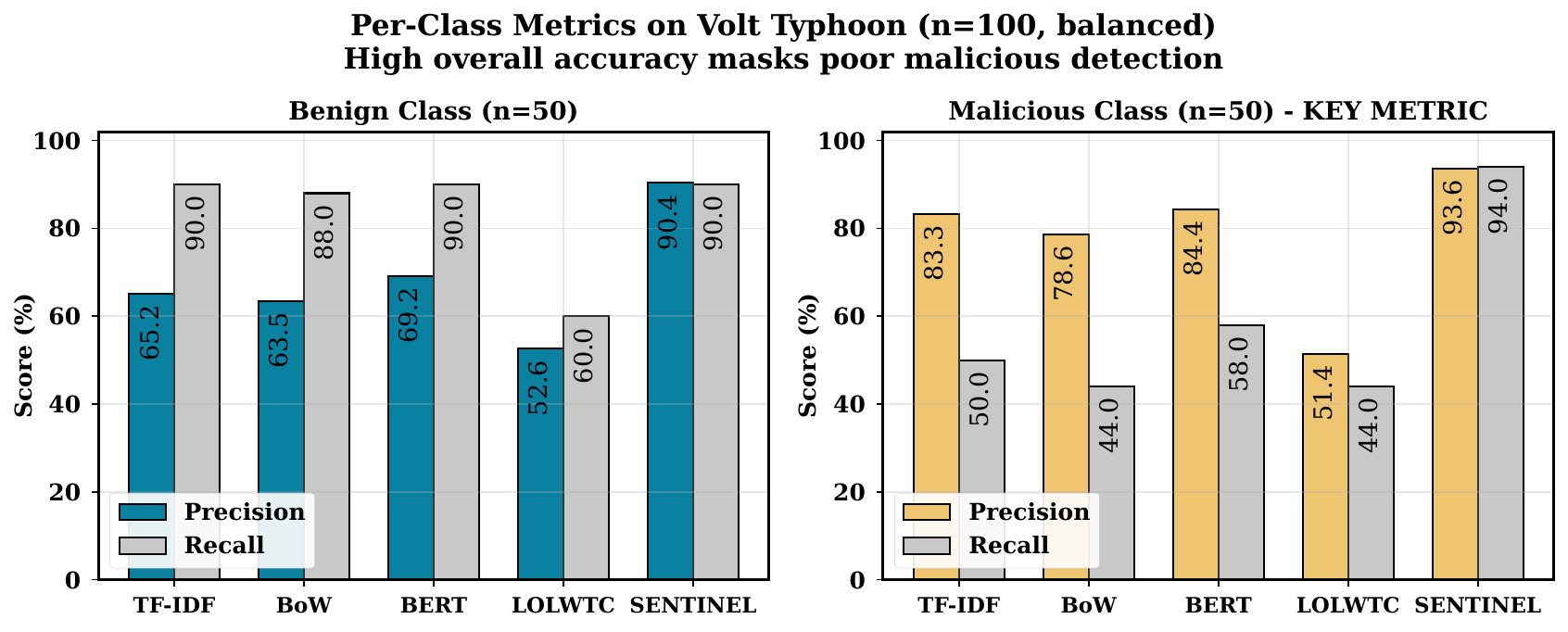}
\vspace{-2em}
\caption{Per-class performance comparison on Volt Typhoon, highlighting stronger malicious detection by SENTINEL.}
\label{fig:perclass}
\end{figure}

Fig.~\ref{fig:perclass} displays per-class precision and recall bars for each model on the balanced Volt Typhoon benchmark, directly exposing the fundamental detection failure hidden by overall accuracy. TF-IDF achieves 98.72\% overall validation accuracy but only 50\% malicious recall on the adversarial set, missing half of documented state-sponsored attack commands. BERT achieves 99.29\% validation accuracy but 58\% malicious recall. This failure arises because models trained on 15:1 imbalanced data converge to classify the dominant benign class accurately while sacrificing sensitivity to the rare malicious class. The consequence for defense operations is operationally unacceptable: a system that misses 42--56\% of state-sponsored commands provides inadequate network protection regardless of its headline accuracy figure.

SENTINEL maintains balanced performance: 90\% benign recall and 94\% malicious recall on the Volt Typhoon set. For defense network monitoring, where missing a coordinated state-sponsored command sequence carries higher operational cost than a false alarm, malicious recall is the primary metric of operational consequence.

\subsection{Ablation Study}

\begin{figure}[!htb]
\centering
\vspace{-0.8em}
\includegraphics[width=0.99\columnwidth,keepaspectratio]{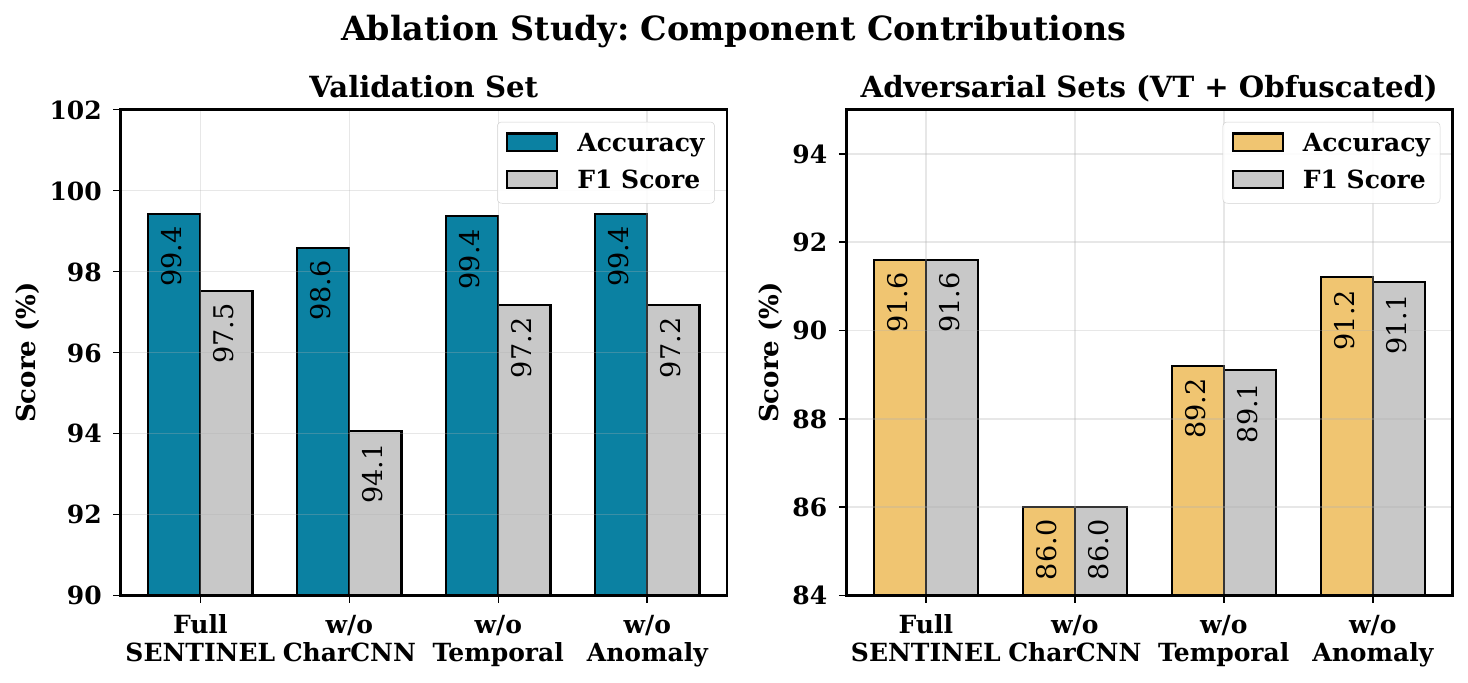}
\vspace{-2em}
\caption{ Ablation study summarizing component contributions on the validation and adversarial evaluation sets. The CharCNN pathway produces the largest observed degradation under adversarial evaluation.}
\label{fig:ablation}
\end{figure}

Fig.~\ref{fig:ablation} shows grouped bars comparing full SENTINEL against three leave-one-out variants across all three datasets, quantifying each pathway's independent contribution. Removing CharCNN degrades Volt Typhoon and obfuscated accuracy by 5.6 percentage points (91.6\% to 86.0\%), validating its role in syntactic invariance (McNemar's $p < 0.001$, $n_\text{eff} = 50$). Disabling inter-command attention reduces Volt Typhoon accuracy by 2.0 percentage points, reflecting value for multi-stage pattern recognition despite training on only 355 malicious sequences. Removing anomaly detection has minimal classification impact (0.4--0.6\%), confirming its role as operational triage for novel techniques rather than primary detection.

The CharCNN contribution is the largest single-pathway gain, confirming that character-level obfuscation invariance is the critical capability missing from all baseline architectures. The inter-command attention contribution is smaller in absolute accuracy terms but addresses a qualitatively distinct failure mode: recognizing coordinated attack progressions that no single-command classifier can detect by design.

\subsection{Low False-Positive Rate Performance}

\begin{figure}[!htb]
\centering
\includegraphics[width=0.99\columnwidth,keepaspectratio]{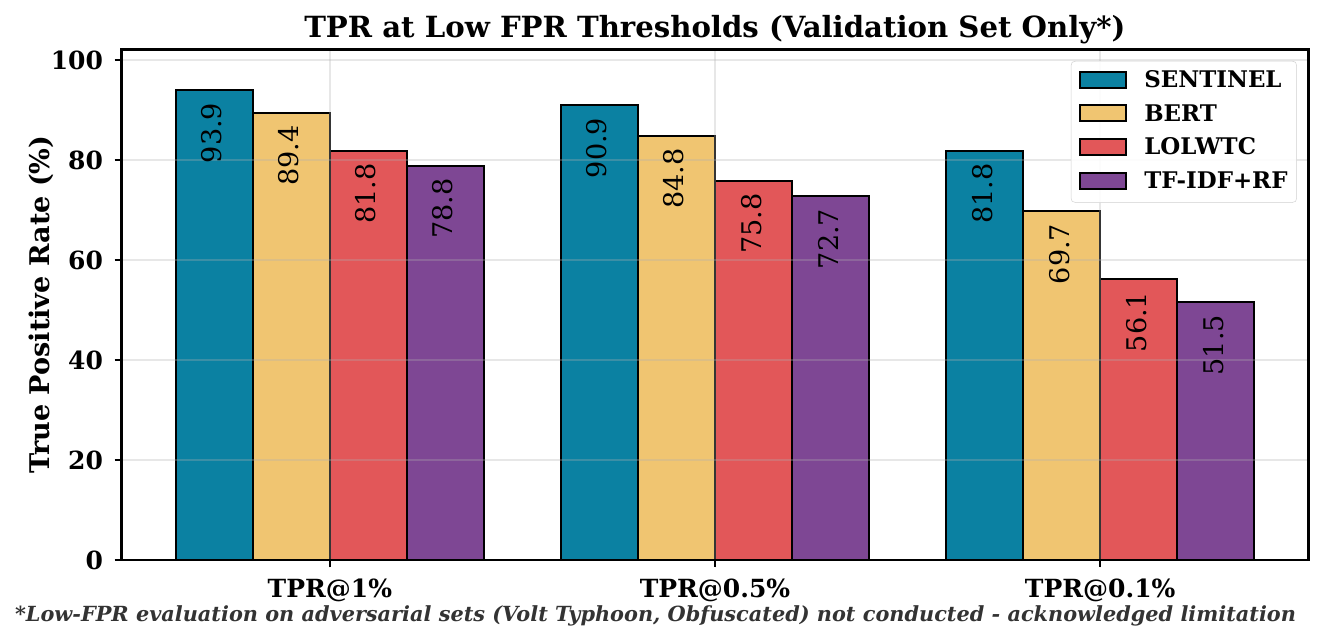}
\vspace{-2em}
\caption{True positive rate (TPR) versus false positive rate (FPR) curves on the validation set show that SENTINEL consistently outperforms baselines at low-FPR operating points, which are critical for SOC deployment. At 0.1\% FPR, SENTINEL achieves 81.8\% TPR, compared to 69.7\% for BERT and 56.1\% for LOLWTC.}
\label{fig:lowfpr}
\end{figure}

Production security operations center deployment requires low false-positive rates to prevent alert fatigue from overwhelming analysts~\cite{liu2022rapid}. Fig.~\ref{fig:lowfpr} compares TPR at three low-FPR operating points; the curves show that SENTINEL maintains a higher true positive rate at every low-FPR operating point on the validation set. At FPR $= 0.1\%$ (one false alarm per 1,000 benign commands), SENTINEL maintains 81.8\% TPR versus 69.7\% for BERT, a 12.1 percentage point improvement. This gain reflects CharCNN's ability to maintain high-confidence classification scores on obfuscated variants that fragment BERT's tokenization into diffuse probability distributions near the decision boundary.

\subsection{Limitations}

\textbf{Platform Scope.} This work addresses Windows command-line LOTL exclusively; Linux-based LOTL using \texttt{curl}, \texttt{python3}, and shell one-liners presents different challenges for which no annotated benchmark at LOLBAS fidelity yet exists, leaving cross-platform extension to future work.

\textbf{Dataset Scale.} With an effective Volt Typhoon test size of $n_{\text{eff}} \approx 50$ after accounting for template dependencies, point estimates carry wide 95\% confidence intervals ([84.8\%, 96.5\%]), and 24\% of malicious samples share semantic patterns with LOLBAS training data, potentially inflating generalization estimates. Performance on enterprise-scale telemetry remains unvalidated.

\textbf{Temporal Scope.} The inter-command attention module operates within fixed 1-hour windows, capturing tactical progressions only; strategic phases separated by days or weeks, as in Volt Typhoon, fall outside this scope and would require hierarchical attention or external memory.

\textbf{Anomaly Calibration.} The 95th-percentile threshold ($\tau = 0.11$) flags roughly 5\% of benign commands, translating to thousands of daily triage alerts at enterprise scale without recalibration. Anomaly scoring should therefore be deployed as an optional analyst-triage tier with independent FPR/TPR calibration.

\section{Conclusion}
We present SENTINEL, an end-to-end framework for detecting Windows LOTL attacks that jointly addresses obfuscation, semantic discrimination, multi-stage temporal reasoning, and distribution shift. Evaluations on the Volt Typhoon dataset highlight that standard benchmark setups can mask critical weaknesses in baseline models, despite apparently strong overall accuracy. Ablation studies further show that each component targets a distinct failure mode, with the full architecture consistently outperforming all individual variants and ablated configurations. These results suggest that robust LOTL detection requires explicitly modeling both syntactic and temporal structure rather than relying on single-view representations. Future work includes extending the framework to Linux LOTL via GTFOBins-based behaviors, improving campaign-level temporal modeling for long-horizon attack tracking, and incorporating richer system and process context signals to support more reliable SOC deployment

\section*{Acknowledgment}

This work is supported by the U.S. Department of Energy(DOE), under Award number DE-NA0004189.

\balance


\end{document}